\documentclass[preprint,aps,prb,showpacs,preprintnumbers,amsmath,amssymb]{revtex4}
\usepackage{graphicx}
\usepackage{amsmath,amssymb}

\begin{document}

\title{Photo-excited semiconductor superlattices as constrained excitable media:
Motion of dipole domains and current self-oscillations}

\author{J. I. Arana}
\author{L. L. Bonilla}\email{bonilla@ing.uc3m.es}

\affiliation{G.\ Mill\'an Institute of Fluid Dynamics, Nanoscience and Industrial
Mathematics, University Carlos III de Madrid, Av.\ Universidad 30, 28911 Legan\'es,
Spain}

\author{H. T. Grahn}
\affiliation{Paul Drude Institute for Solid State Electronics,
Hausvogteiplatz 5--7, 10117 Berlin, Germany}

\date{\today}

\begin{abstract}
A model for charge transport in undoped, photo-excited semiconductor
superlattices, which includes the dependence of the electron-hole
recombination on the electric field and on the photo-excitation
intensity through the field-dependent recombination coefficient,
is proposed and analyzed.
Under dc voltage bias and high photo-excitation intensities, there appear
self-sustained oscillations of the current due to a repeated
homogeneous nucleation of a number of charge dipole waves inside
the superlattice. In contrast to the case of a constant recombination
coefficient, nucleated dipole waves can split for a field-dependent
recombination coefficient in two oppositely moving dipoles.
The key for understanding these unusual properties is that
these superlattices have a unique static electric-field domain.
At the same time, their dynamical behavior is akin to the one of an
extended excitable system: an appropriate finite disturbance of the
unique stable fixed point may cause a large excursion in phase space
before returning to the stable state and trigger pulses and wave trains.
The voltage bias constraint causes new waves to be nucleated when old
ones reach the contact.
\end{abstract}

\maketitle

\section{Introduction}
\label{sec:intro}
Nonlinear charge transport in weakly coupled, undoped, photo-excited, type-I
semiconductor superlattices (SLs) is well described by spatially discrete
drift-diffusion equations.\cite{Bon94,Bon95}
As in the much better known case of doped SLs, nonlinear phenomena
include formation and dynamics of electric-field domains, self-sustained oscillations of
the current through voltage-biased SLs, chaos, etc. Experimentally, the formation of
static electric-field domains in undoped, photo-excited SLs was already reported many years
ago.\cite{Gra90} The first experimental observation of dynamical aspects of domain
formation in undoped, photo-excited SLs such as self-sustained oscillations of the 
photo-current were reported by Kwok \textit{et al.}\cite{Kwo95}
Due to the excitation condition, the oscillations were damped.
Subsequently, undamped self-sustained oscillations of the photo-current in undoped SLs were
observed for a type-II GaAs/AlAs\cite{Hos96} and for a direct-gap GaAs/AlAs SL.\cite{Oht97}
Tomlinson \textit{et al.}\cite{Tom99} reported the detection of undamped photo-current
oscillations in an undoped GaAs/Al$_{0.3}$Ga$_{0.7}$As SL, where the transport is governed
by resonant tunnelling between $\Gamma$ states. The evolution from a static state at
low carrier densities to an oscillating state at higher carrier densities was demonstrated in an 
undoped, photo-excited SL by increasing the photo-excitation intensity.\cite{Luo99}
An investigation of the bifurcation diagrams for undoped, photo-excited SLs
showed the existence of a transition between periodic and chaotic oscillations.\cite{Oht98}
For a detailed review of the nonlinear static and dynamical properties of doped and
undoped superlattices, see Ref.~\onlinecite{Bon05}.

Previous theoretical studies of undoped photo-excited SLs, including studies of
bifurcation and phase diagrams,\cite{Per01} are based on a discrete drift model
having a constant recombination coefficient.\cite{Bon94} So far, there have been
no reports on considering field-dependent recombination or the fact that the
time scale of the electron-hole dynamics depends strongly on the optical
excitation intensity. However, the consequences of including these effects for
the dynamics of electric-field domains can be striking. In this work, we incorporate
into the previously studied discrete model the dependence of the
electron-hole recombination on the electric field and on the photo-excitation intensity
using a straightforward model that takes into account the overlap integral between
the electron and hole wave functions. The field-dependent recombination coefficient
decreases with increasing electric field, which has far reaching consequences.

At high photo-excitation intensities, it is possible to find only one stable
electric-field domain, not two as in the case of a constant recombination
coefficient.\cite{Bon94} In this case, self-sustained oscillations of the current (SSOC)
may appear under dc voltage bias. The field profile during SSOC
can exhibit nucleation of dipole waves inside the sample, the splitting of one wave
into two, and the motion of the resulting waves in opposite directions.
These dipole waves resemble the pulses in excitable reaction-diffusion systems such as
the FitzHugh-Nagumo model for nerve
conduction \cite{Kee98,Car03,Car05,Bon09} and are quite different from field profiles
for a constant recombination coefficient.\cite{Bon94,Bon95,Bon05}
In an excitable dynamical system, an appropriate finite disturbance of the unique
stable fixed point may cause a large excursion in phase space before returning to the
stable state. When diffusion is added, the resulting reaction-diffusion
system may support wave fronts, pulses, and wave trains. It is also possible
to find SL configurations at high photo-excitation intensities for which there exist
no stable electric-field domains. In these cases, there are SSOC, whose 
corresponding field profiles are wave trains, comprising a
periodic succession of dipole waves. These cases are similar to wave trains in
oscillatory media such as those appearing in the FitzHugh-Nagumo model in the presence
of a sufficiently large external current.\cite{Kee98,Bon09}

\section{Model equations}
\label{sec:model}
\begin{figure}[b!]
\begin{center}\leavevmode
\includegraphics[width=0.5\linewidth]{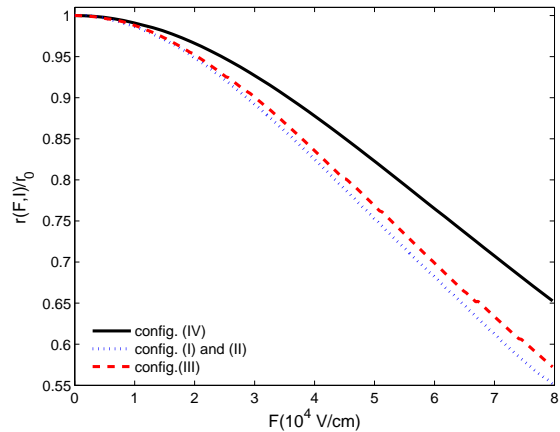}
\caption{(Color online) Recombination coefficient $r(F,I)/r(0,I)$ vs electric field $F$ for the four configurations
listed in Table \ref{t1}.}
\label{fig1}
\end{center}
\end{figure}

The equations governing nonlinear charge transport in weakly coupled, undoped,
photo-excited, type-I SLs are
\begin{eqnarray}
&&\varepsilon\,  (F_{i}-F_{i-1})= e\, ( n_{i} -p_{i})\;,\quad i=1,\ldots, N,
\label{1}\\
&& \varepsilon\, \frac{dF_{i}}{dt} + _{i\rightarrow i+1} = J(t)\;, \quad i=0,1,\ldots, N,\label{2}\\
&&\frac{dp_{i}}{dt}=\gamma(I) - r(F_{i},I)\, n_{i}p_{i}\;, \quad i=1,\ldots, N.
\label{3}
\end{eqnarray}
In these equations, the tunneling current densities between the quantum wells (QWs)
as well as between the SL and the contact regions are
\begin{eqnarray}
&& J_{i\rightarrow i+1} =\frac{en_{i} v(F_{i})}{l}-eD_{i}(F_{i})\frac{n_{i+1}
-n_{i}}{l^{2}}\;,  \label{4} \\
&& J_{0\rightarrow 1} = \sigma\, F_{0}\;,\quad\quad
J_{N\rightarrow N+1} =\frac{n_{N}}{ N_{D}}\sigma\, F_{N}\;. \label{5}
\end{eqnarray}
The voltage bias condition is
\begin{eqnarray}
&&\frac{1}{N+1}\sum_{i=0}^{N} F_{i} = \phi \equiv\frac{V}{l\, (N+1)}\;.
\label{6}
\end{eqnarray}
Here $-e < 0$, $\varepsilon$, $\sigma$, $N_D$, $-F_{i}$, $n_{i}$ and $p_{i}$ denote the electron
charge, the average permittivity, the conductivity of the injecting contact, the doping density of the
collecting contact, the average electric field, as well as the two-dimensional electron and
hole densities of the $i$th period of the SL, respectively. Equation~(\ref{1}) corresponds
to the averaged Poisson equation. Equation~(\ref{2}) denotes Amp\`ere's law: the total
current density $J(t)$ equals the sum of the displacement current density and
$J_{i\to i+1}$, the electron tunneling current density across the $i$th barrier that
separates the quantum wells $i$ and $i+1$. Charge continuity is obtained by
differentiating Eq.~(\ref{1}) with respect to time and using Eq.~(\ref{2}) in the result.
Tunneling of holes is neglected so that only photogeneration and recombination of holes with
electrons enter into Eq.~(\ref{3}). For high temperatures, i.e. $k_B T\gg \pi\hbar^2[n]/m^*$,
where $k_B$ denotes Boltzmann's constant, $T$ the temperature, $m^*$ the
effective electron mass, and $[n]$ the order of magnitude of the electron density,
the tunneling current is given by Eq.~(\ref{4}), in which $v(F_{i})$ and $D(F_{i})$ are functions of the
electric field given in Ref.~\onlinecite{Bon05} and $l=L_{w}+L_{B}$ is the length of one SL period
($L_{w}$ and $L_{B}$ denote the individual widths of the quantum well and barrier, respectively).
Even for lower temperatures, the qualitative behavior of the solutions of the discrete drift-diffusion
model is similar to one of the more general tunneling current models described in
Ref.~\onlinecite{Bon05}. For $D=0$ and a constant recombination coefficient $r$,
Eqs.~(\ref{1})--(\ref{4}) describe the well known discrete drift model introduced in
Ref.~\onlinecite{Bon94}. The indices $i = 0$ and $i = N + 1$ represent the SL injecting
and collecting contacts, and Eq.~(\ref{2}) holds for them with the phenomenological
currents given by Eq.~(\ref{5}). The total current density follows from the voltage bias
condition in Eq.~(\ref{6})
\begin{eqnarray}
J(t)= \frac{1}{N+1}\sum_{i=0}^{N} J_{i\to i+1} + \varepsilon\,\frac{d\phi}{dt}\;.
\label{7}
\end{eqnarray}
Photogeneration and recombination are given by
$\gamma(I) = I\,\alpha_{\rm 3D}(\hbar\omega_{\rm exc})\, L_{w}/(\hbar
\omega_{\rm exc})$ and
\begin{equation}
r(F,I)= \left(\frac{n_{\rm ref}}{n_{\rm in}\pi c}\right)^2\int_{0}^\infty
\frac{\omega^2\alpha_{\rm 2D}(\hbar\omega, F)}{\exp(\frac{\hbar\omega}{k_{B}
T})-1}\, d\omega\;,
\label{8}
\end{equation}
respectively. Here $I$, $\omega_{\rm exc}$, $n_{\rm ref}$, $n_{\rm in}\approx\gamma
n_{\rm ref} L_{w}/c$, and $c$ denote the photo-excitation intensity, the frequency
of the exciting photon, the refractive index, the intrinsic carrier density, and the
speed of light, respectively. $\alpha_{\rm 2D}$ and $\alpha_{\rm 3D}$
correspond to the two- (2D) and three-dimensional (3D) absorption coefficients. The 2D
absorption coefficient is proportional to the square of the modulus of the electron-hole
overlap integral for a constant electric field $F$ (cf. Ref.~\onlinecite{Gra99})
\begin{eqnarray}
\alpha_{2D}(\omega,F) &=&\alpha _{0}^{2D}\int_{0}^{\infty}\delta(
E_{k_{\Vert }}+E_{g}-\hbar\omega +\tilde{E}_{e}^{n}(F)+
\tilde{E}_{h}^{m}(F))\, d\tilde{E}_{k_{\Vert }}  \label{9} \\
&&\times \left\vert \int_{-l/2}^{l/2}\Psi_{e}^{n}(\tilde{z},F)\Psi_{hh}^{m}(
\tilde{z},F)d\tilde{z}\right\vert ^{2}\;,
\notag
\end{eqnarray}
where $E_{g}$ denotes the energy of the bandgap at the $\Gamma$ point and $\Psi_{n}$ as well
as $\Psi_{h}$ solve the stationary Schr\"odinger equation inside one SL period, $[-l/2,l/2]$,
for the electrons and holes, respectively. In this equation, the electric field $F$ is considered
to be constant and $\Psi_{n,h}(\pm (L_{w}/2+l_{p\pm}))=0$, where the penetration length $l_{p\pm}$
solves the cubic equation
\begin{equation}
l_{p\pm}\,\sqrt{2m^*\left[V-eF\left(l_{p\pm}\mp\frac{L_{w}}{2}\right)-E_{n,h}
\right]}= 1\;, \nonumber
\end{equation}
and therefore $l_{p\pm}$ depends self-consistently on the eigenvalue
$E_{n,h}$. For a fixed value of $I$, the recombination coefficient decreases with increasing
electric-field strength $F$, as depicted in Fig.~\ref{fig1}.

\section{Dimensionless equations}
\label{sec:nondim}
To study the model described by Eqs.~(\ref{1})--(\ref{7}), it is convenient to render them
dimensionless. For a fixed value of the photo-excitation intensity $I$ and a constant electric field $F_{i}=F$,
the stationary solution of Eq.~(\ref{3}) is $n_{i}=p_{i}=\sqrt{\gamma(I)/r(F,I)}$. We use the
maximum values $[n]=[p]=\sqrt{\gamma(I)/r(0,I)}$ to define typical values for $n_{i}$ and
$p_{i}$. The field $F_{M}$ at the maximum of the drift velocity is a typical value of the
field when there are SSOC. Therefore, we adopt it as the field unit $[F]=F_{M}$.
Similarly, $[v]=v_{M}$ and, therefore, $[J]=e[n]v_{M}/l$ and $[D]=lv_{M}$. There are two
possible time scales, the first one being $t_{F}= \varepsilon F_M/[J]$, which balances Maxwell's displacement
current with the current density in Eq.~(\ref{2}), and the second one $t_{n}=[p]/\gamma(I)=1/\sqrt{\gamma(I)
r(0,I)}$, which balances both sides of Eq.~(\ref{3}). It is reasonable to choose the time unit as
the longer of the two times $t_{F}$ and $t_{n}$. We have chosen four representative
GaAs/Al$_{x}$Ga$_{1-x}$As SL configurations with 10~nm wells and 4~nm barriers. The configurations
are I: $x=0.25$ and $I= 120.5$ kW/cm$^2$; II: $x=0.25$ and $I=302.69$ kW/cm$^2$; III: $x=0.3$ and
$I=479.735$ kW/cm$^2$; and IV: $x=1.0$ and $I=30.27$ kW/cm$^2$.
We assume a circular cross section with a diameter of 160~$\mu$m, a photo-excitation
intensity of 60 mW, a wavelength of 413~nm, and four different beam diameters yielding the
previously listed values of the laser intensity. For most of the cases listed in Table \ref{t1},
$t_{n}>t_{F}$, and therefore we choose $[t]=t_{n}$. All scaled variables are listed in Table \ref{t1}.
\begin{table}[b!]
\caption{Units used to achieve a set of equations with dimensionless variables.}
\label{t1}
\begin{center}\begin{tabular}{ccccccccccc}
\hline
config. & $x$& $I$& $n$, $p$ & $F$, $\phi$ &$t$& $v$ &$J$& $D$& $r$& $\sigma$ \\
\hline
&  & & $\sqrt{\frac{\gamma(I)}{r(0,I)}}$ & $F_{M}$ & $\frac{[n]}{\gamma(I)}$ & $v_{M}$
& $\frac{e[n]v_{M}}{l}$& $lv_{M}$ & $r(0,I)$& $\frac{ev_{M}[n]}{F_{M}l}$\\
& &$\frac{\mbox{kW}}{\mbox{cm}^2}$& $\frac{10^{12}}{\mbox{cm}^{2}}$ &$\frac{\mbox{kV}}{\mbox{cm}}$ & $10^{-11}$
s & $\frac{\mbox{km}}{\mbox{s}}$ & $10^4\frac{\mbox{A}}{\mbox{cm}^2}$ &
$10^{-2}\frac{\mbox{cm}^2}{\mbox{s}}$&
$10^{-3}\frac{\mbox{cm}^2}{\mbox{s}}$& $\frac{\mbox{A}}{\mbox{V cm}}$ \\
I & 0.25& 120.5& 1.017 & 16.8 & 4.602 &2.569 & 2.991& 35.97 & 5.7 & 1.78\\
II & 0.25 & 302.69& 4.0504 & 14.72 &18.321 & 0.0057 & 11.910 & 35.97 & 0.90026 & 7.0891 \\
III & 0.3& 479.735 & 8.070 & 16.0 & 69.485 &1.421 & 13.12&19.89& 0.359&8.912\\
IV 
& 1& 30.27&0.2731 &14.72 & 313.45 &0.0057& 8.28$\times 10^{-4}$&7.95$\times 10^{-4}$&
90.7&5.62$\times 10^{-4}$\\
\hline
\end{tabular}
\end{center}
\end{table}

We now rewrite the model equations using dimensionless variables by defining $\hat{n}_{i}= n_{i}/[n]$,
$\hat{t}=t/[t]$, \ldots, where $[n]$,  $[t]$, etc.\ are the scales defined above and specified
in Table \ref{t1}. Omitting hats over the variables,
the dimensionless system of equations corresponding to Eqs.~(\ref{1})--(\ref{7}) read
\begin{eqnarray}
&&F_{i}-F_{i-1}=(n_{i} -p_{i})\nu\;, \label{10}\\
&& \delta\,\frac{dF_{i}}{dt} +n_{i}v(F_{i}) - D(F_{i})\, (n_{i+1}-n_{i}) = J(t)\;,
\label{11}\\
&&\frac{dp_{i}}{dt}=1 - r(F_{i})\, n_{i}p_{i}\;,\label{12}\\
&& \sigma\, F_{0}+\delta\,\frac{dF_{0}}{dt}=J\;,\quad\quad
\sigma\alpha\, n_{N}F_{N}+\delta\,\frac{dF_{N}}{dt}=J\;, \label{13}
\end{eqnarray}
\begin{eqnarray}
&&\frac{1}{N+1}\sum_{i=0}^{N} F_{i} = \phi\;,
\label{14}\\
&& J = \frac{1}{N+1}\sum_{i=0}^{N} J_{i\to i+1} + \frac{d\phi}{dt}\;.
\label{15}
\end{eqnarray}
In these equations, there are four dimensionless parameters,
\begin{eqnarray}
\delta=\frac{\varepsilon F_{M}l r(0,I)}{e v_{M}},\quad \nu= \frac{e[n]}
{\varepsilon F_{M}}, \quad \alpha=\frac{[n]}{N_{D}}\equiv
\frac{\sqrt{\gamma(I)}}{N_{D}\sqrt{r(0,I)}}\;, \label{16}
\end{eqnarray}
and the dimensionless conductivity $\hat{\sigma}$, which is here simply denoted by $\sigma$.
The values of the dimensionless parameters for the four SL configurations given in Table~\ref{t1} are
listed in Table \ref{t2}.
It is interesting to note that $r(0,I)\propto I^{-2}$ and $\gamma(I)\propto I$ so that
$\nu\propto I^{3/2}$ and $\delta\propto I^{-2}$, i.e. $\nu$ increases with photo-excitation
intensity, whereas $\delta$ decreases. High photo-excitation intensities imply that $\nu$ is large and
$\delta$ small, whereas the opposite holds for low photo-excitation intensities.
\begin{table}[b!]
\caption{Numerical values of the dimensionless parameters $\alpha$, $\delta$, $\nu$
and $\hat{\sigma}$ for the four superlattice configurations listed in Table~\ref{t1}.}
\label{t2}
\begin{center}\begin{tabular}{ccccc}
 \hline
config. & $\alpha$ & $\delta$ & $\nu$ & $\hat{\sigma}$ \\
&  $\frac{[n]}{N_{D}}$ & $\frac{\varepsilon F_M l\, r(0,I)}{e v_M}$ &
$\frac{e[n]}{\varepsilon F_{M}}$& $\frac{\sigma F_{M}l}{ev_{M}[n]}$ \\
\hline
I   & 0.9713 & 0.0037 & 8.446& 1.0834\\
II  &0.6944 & $5.91\times 10^{-4}$& 33.6228& 1.0523\\
III & 0.9779& $4.05\times 10^{-5}$ & 70.507& 1.1074 \\
IV & 1.5754 & 22.3714& 1.2721& 1.2959\\
\hline
\end{tabular}
\end{center}
\end{table}

\begin{figure}[b!]
\begin{center}\leavevmode
\includegraphics[width=0.5\linewidth]{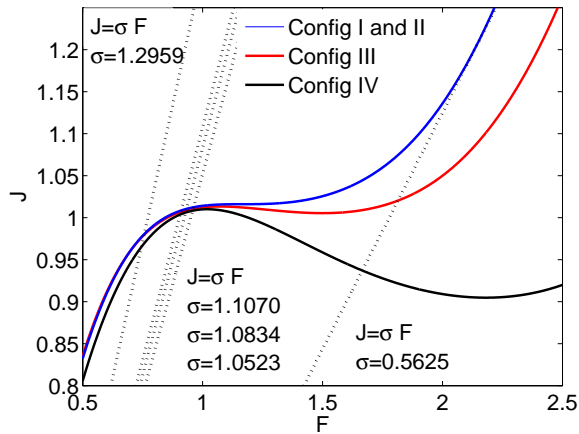}
\caption{(Color online) 
Local current density $j$ vs electric field $F$. For a fixed value of the total current
density $J$, there may be one or three zeros of $j(F)$-$J$ depending on the Al content $x$.}
\label{fig2}
\end{center}
\end{figure}
It is interesting to depict the phase plane corresponding to spatially uniform solutions of
Eqs.~(\ref{10})--(\ref{12}) with $n_{i}=p_{i}=p$, $F_{i}=F$
\begin{eqnarray}
\delta\,\frac{dF}{dt} =J-p\, v(F), \quad \frac{dp}{dt}=1 - r(F)\, p^2\;.\label{17}
\end{eqnarray}
Generically and for a fixed value of $J$, the nullclines $v(F)\, p=J$ and $r(F)\, p^2=1$ intersect in
one or three fixed points, depending on the Al content $x$ in the barriers. At these
fixed points,
\begin{equation}
j(F) = J, \quad j(F)=\frac{v(F)}{\sqrt{r(F)}}\;.
\label{j-f}
\end{equation}
The function $j(F)$ is depicted in Fig.~\ref{fig2} for the SL configurations
listed in Table \ref{t1}. 

The calculations for arbitrary values of Aluminum content $x$ show that for
$0.45\leq x\leq 1$, there are three fixed
points of the system in Eq. (\ref{17}), one on each of the three branches of $p=J/v(F)$,
and the ratio of $(j_{\rm max}-j_{\rm min})$ to the average current $(j_{\rm max}+
j_{\rm min})/2$ is sufficiently large. As we shall see later, some nonlinear phenomena
occurring in these SLs are quite similar to the ones observed in doped SLs: static electric-field
domains with domain walls joining the stable branches of $p=J/v(F)$, SSOC
due to the recycling of pulses formed by two moving domain walls having a
high-field region between them, etc. For $0<x<0.45$, $j(F)$ is either increasing
for positive $F$ (for $0<x<0.25$) or  the ratio of $(j_{\rm max}-j_{\rm min})$ to
the average current $(j_{\rm max}+j_{\rm min})/2$ is small (for $0.25<x<0.45$). For
$0<x<0.25$, there is a unique fixed point at $F=F_{*}$, which, for an appropriate value of
$J$, may be
located on any of the three branches of $p=J/v(F)$. If the fixed point is located on one
of the two stable branches of $p=J/v(F)$ for which $v(F)$ has a positive slope,
the dynamical system of Eq.~(\ref{17}) is excitable, whereas it is oscillatory if the fixed point is located
on the second branch of $p=J/v(F)$ with $v'(F_*)<-2\delta r(F_*)<0$. In an excitable dynamical system, an
appropriate finite disturbance of the unique stable fixed point may cause a large excursion in phase
space before returning to the stable state. When diffusion is added, the resulting excitable
reaction-diffusion system may support a variety of wave fronts, pulses and wave
trains.\cite{Kee98,Car03,Car05,Bon09} A dynamical system having an unstable fixed point and a stable
limit cycle around it is oscillatory. Again in the presence of diffusion, oscillatory systems
may support different spatio-temporal patterns.\cite{Kee98,Bon09,Kur03} An undoped photo-excited
SL with excitable or oscillatory dynamics exhibits quite unusual phenomena. Under dc current
bias, it is possible to have pulses moving to the right or to the left and periodic wave
trains. Under dc voltage bias, these pulses and wave trains may give rise to a variety of
SSCOs. Similar phenomena are observed in the case $0.25<x<0.45$ for which
$j(F)$ has a shallow valley for a narrow interval of current densities.

\section{dc voltage-biased superlattice for small photo-excitation intensities}
\label{sec:voltage}

The behavior of a dc voltage-biased SL is quite different depending on its Al
content and photo-excitation intensity. For an Al content smaller than
45\%, $j(F)$-$J$ in Eq.~(\ref{j-f}) has a single zero for any value of $J$, and
the only stable states of the SL are stationary ones unless the photo-excitation
intensity is sufficiently large (cf. next section). Let us assume that the Al content
is larger than 45\% so that $j(F)$-$J$ in Eq.~(\ref{j-f}) may have three zeros for an
appropriate range of $J$ values. In this case, the undoped SL behaves similarly to an $n$-doped SL.\cite{Bon05}
The most interesting limit is that of small photo-excitation intensity, i.~e., $\delta\gg 1$.

\begin{figure}[b!]
\begin{center}\leavevmode
\includegraphics[width=0.5\linewidth]{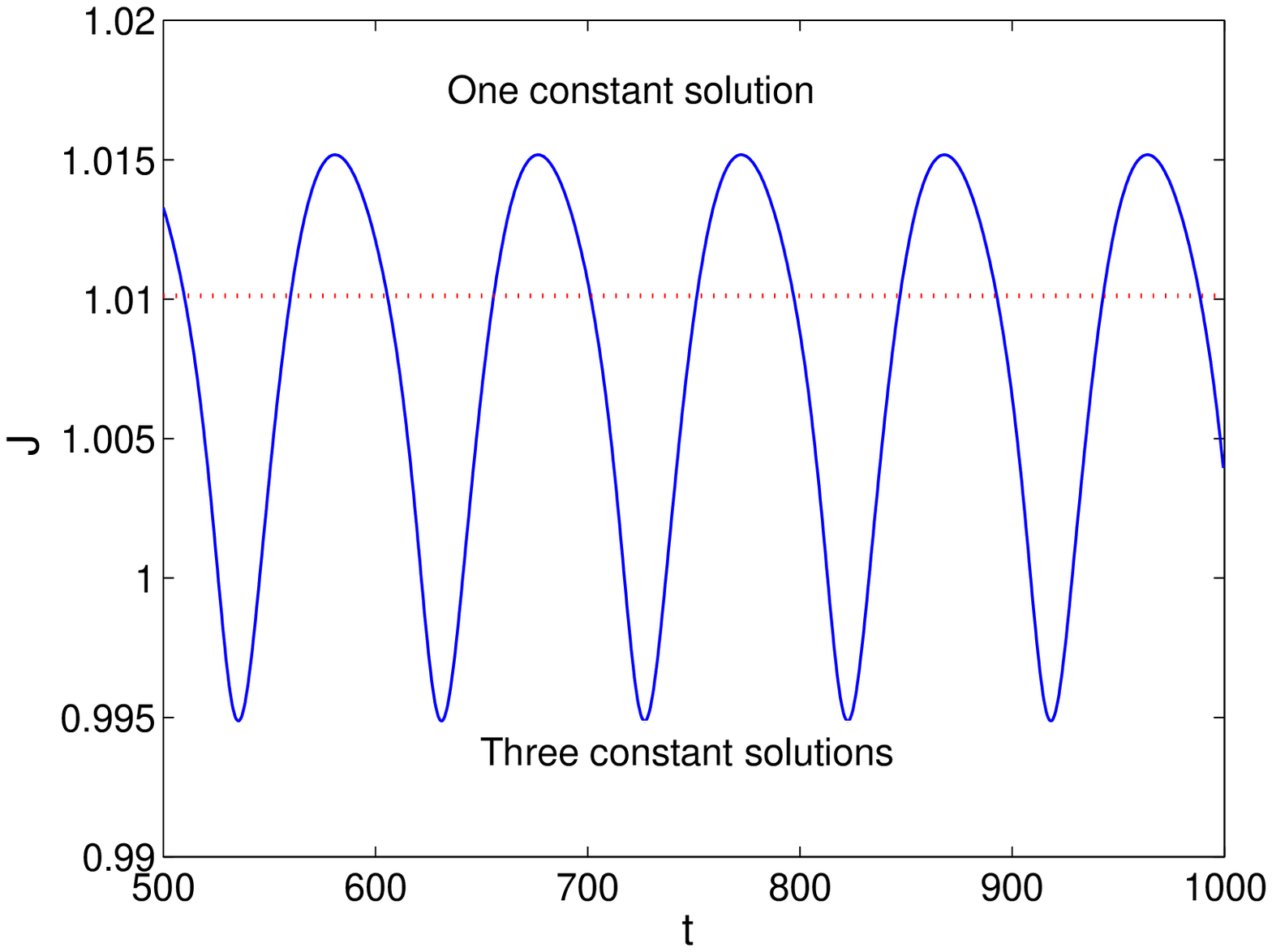}
\includegraphics[width=0.5\linewidth]{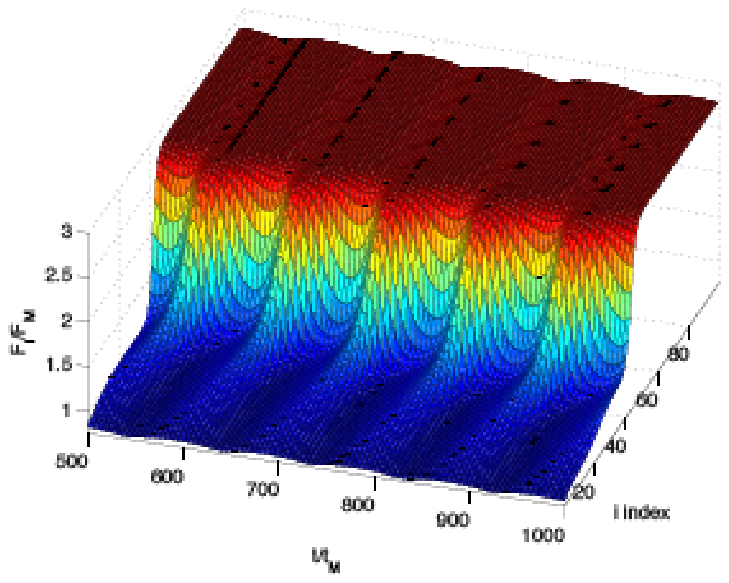}
\caption{(Color online) Current vs time (top) and field distribution vs time (bottom)
displaying SSCOs due to monopole recycling at the injecting contact. The injecting contact
conductivity and the bias are $\sigma=1.2959$ and $V =2.0416236$, respectively. The other
parameter values correspond to configuration IV in Table \ref{t1} and
are $N=99$, $\delta=22.35636$, $\sigma_{2} = 1.3148$, $\nu= 1.27214$, and Al content $x=1$.}
\label{fig3}
\end{center}
\end{figure}
\begin{figure}[t!]
\begin{center}\leavevmode
\includegraphics[width=0.5\linewidth]{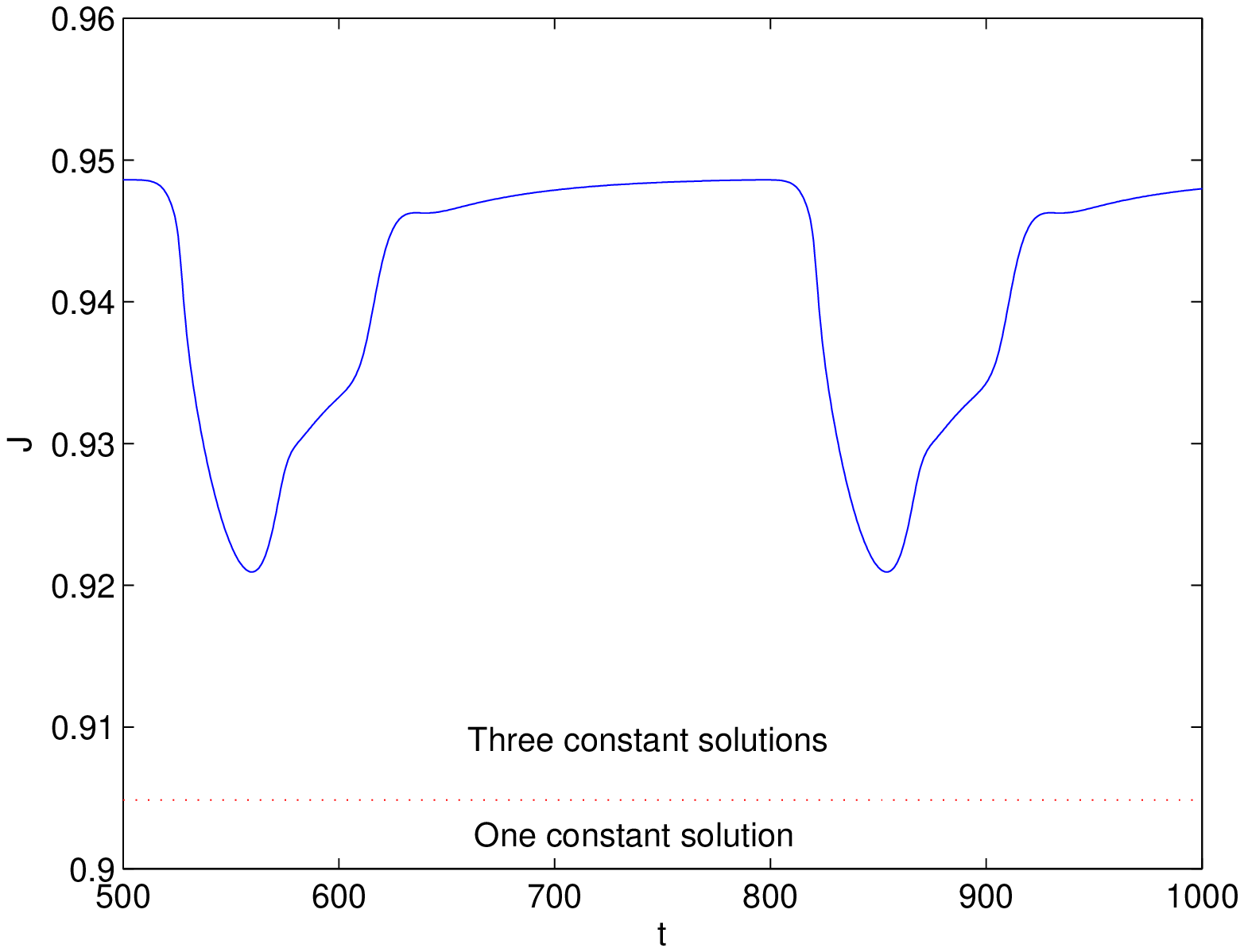}
\includegraphics[width=0.5\linewidth]{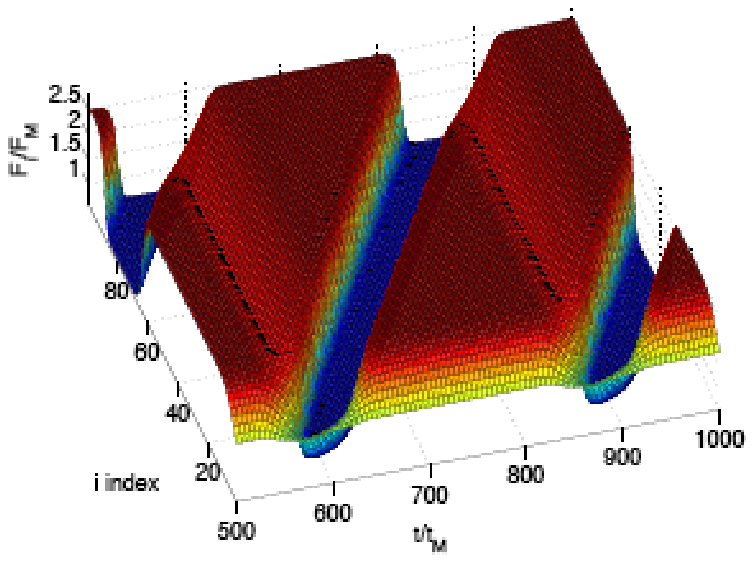}
\caption{(Color online) Current vs time (top) and field distribution vs time (bottom)
displaying SSCOs due to dipole recycling at the injecting contact. The injecting contact
conductivity and the bias are $\sigma=0.5625$ and $V =2.05148375$, respectively. The other parameter
values are the same as the ones used for Fig.~\ref{fig3} corresponding to configuration IV in
Table \ref{t1}. }
\label{fig4}
\end{center}
\end{figure}

For the time scale $\tau=t/\delta$, Eqs.~(\ref{11}) and (\ref{12}) can be rewritten as
\begin{eqnarray}
\frac{dF_{i}}{d\tau} &=& J -\left(p_{i}+\frac{F_{i}-F_{i-1}}{\nu}\right)
v(F_{i}) \nonumber\\
&+& \left(p_{i+1}-p_{i}+\frac{F_{i+1}+F_{i-1}-2F_{i}}{\nu}\right) D(F_{i})\;,
\label{49}\\
\delta^{-1}\frac{dp_{i}}{d\tau}&=& 1-\left(p_{i}+\frac{F_{i}-
F_{i-1}}{\nu}\right)p_{i}\, r(F_{i})\;.\label{50}
\end{eqnarray}
In the limit of large photo-excitation intensities, $\delta\ll 1$, Eq.~(\ref{50}) indicates that the $p_i$ do not
depend on $\tau$. In this case, Eq.~(\ref{49}) corresponds to a SL doped with a density $p_i$, and
we may expect phenomena similar to the ones observed in an $n$-doped SL. In the opposite
limit of small photo-excitation intensities, $\delta^{-1}\ll1$, $\tau=\delta^{-1}t$ is a slow scale. The
$p_{i}$ are functions of $F_{i}-F_{i-1}$ determined by solving Eq.~(\ref{50}) with a zero left hand side
\begin{eqnarray}
p_{i}=\sqrt{\left(\frac{F_{i}-F_{i-1}}{2\nu}\right)^2 + \frac{1}{r(F_{i})}}
-\frac{F_{i}-F_{i-1}}{2\nu}\;.  \label{51}
\end{eqnarray}
The resulting equation for $F_{i}$ is then obtained by inserting Eq.~(\ref{51}) into Eq.~(\ref{49})
\begin{eqnarray}
\frac{dF_{i}}{d\tau} &=& J -\left(\sqrt{\left(\frac{F_{i}-F_{i-1}}{2\nu}
\right)^2 + \frac{1}{r(F_{i})}}+\frac{F_{i}-F_{i-1}}{2\nu}\right)v(F_{i})
\nonumber\\
&+& \left[\sqrt{\left(\frac{F_{i+1}-F_{i}}{2\nu}
\right)^2 + \frac{1}{r(F_{i+1})}}-\sqrt{\left(\frac{F_{i}-F_{i-1}}{2\nu}
\right)^2 + \frac{1}{r(F_{i})}}\right.\nonumber\\
&+&\left.\frac{F_{i+1}+F_{i-1}-2F_{i}}{2\nu}\right]D(F_{i})\;.
\label{52}
\end{eqnarray}
This equation is similar to the one describing the electric field in a doped SL, but now there
are drift and diffusion terms which are nonlinear in the differences $F_{i}-F_{i-1}$.
Under dc voltage bias, there are SSOC mediated by pulses of the
electric field. The current density varies on a slow time scale, whereas the electric-field profile
consists of a varying number of wave fronts joining the stable constant solutions of Eq.~(\ref{52})
at the instantaneous value of the current.\cite{Bon05} Depending on the conductivity of the
injecting contact $\sigma$, there are different types of SSOC due to
the periodic generation of dipoles or monopoles at the injecting contact. If the curve $\sigma F$
intersects the bulk current-field characteristic curve $j(F)$ before its maximum
(cf. Fig.~\ref{fig2}), 
SSCOs due to recycling and motion of charge monopole waves (moving charge
accumulation layers) appear as shown in Fig.~\ref{fig3}. 
If $\sigma F$ intersects $j(F)$ after
its maximum (cf. Fig.~\ref{fig2}), 
SSCOs due to recycling and motion of dipole waves are obtained,
as depicted in Fig.~\ref{fig4}. 
We have indicated in Figs.~\ref{fig3} 
and \ref{fig4} 
whether there are one or three uniform and time-independent (constant) solutions of Eq.~(\ref{17}),
solving $p\, v(F)=J$ and $r(F)\, p^2=1$ for the instantaneous value of the current density $J=J(t)$
during the SSOC.

\section{dc voltage-biased superlattice for large photo-excitation intensities}
\label{sec:smallx}

If $0<x<0.45$, the phase plane in Eq.~(\ref{17}) may have only one fixed point located on any
branch of the nullcline $p=J/v(F)$. For small photo-excitation intensities, the only stable state is
a stationary one. However, for sufficiently large photo-excitation intensities, an infinite, dc current-biased SL
may exhibit pulses moving downstream or upstream  and also wave trains moving downstream.\cite{abg_math}
The counterpart of these stable solutions for a dc voltage-biased SL is very
interesting and different from anything observed in an $n$-doped SL. Our simulations
correspond to SLs with different Al contents, whose current-field characteristics $j(F)$ and
injecting contact curve $j=\sigma F$ are shown in Fig.~\ref{fig2}. 
In all cases, SSCOs appear for an average bias roughly in the region of negative differential resistance
(NDR), where $j'(F)<0$ (e.g., $\sigma$=0.5625 in Fig.~\ref{fig2}). 

\begin{figure}[t!]
\begin{center}\leavevmode
\includegraphics[width=0.5\linewidth]{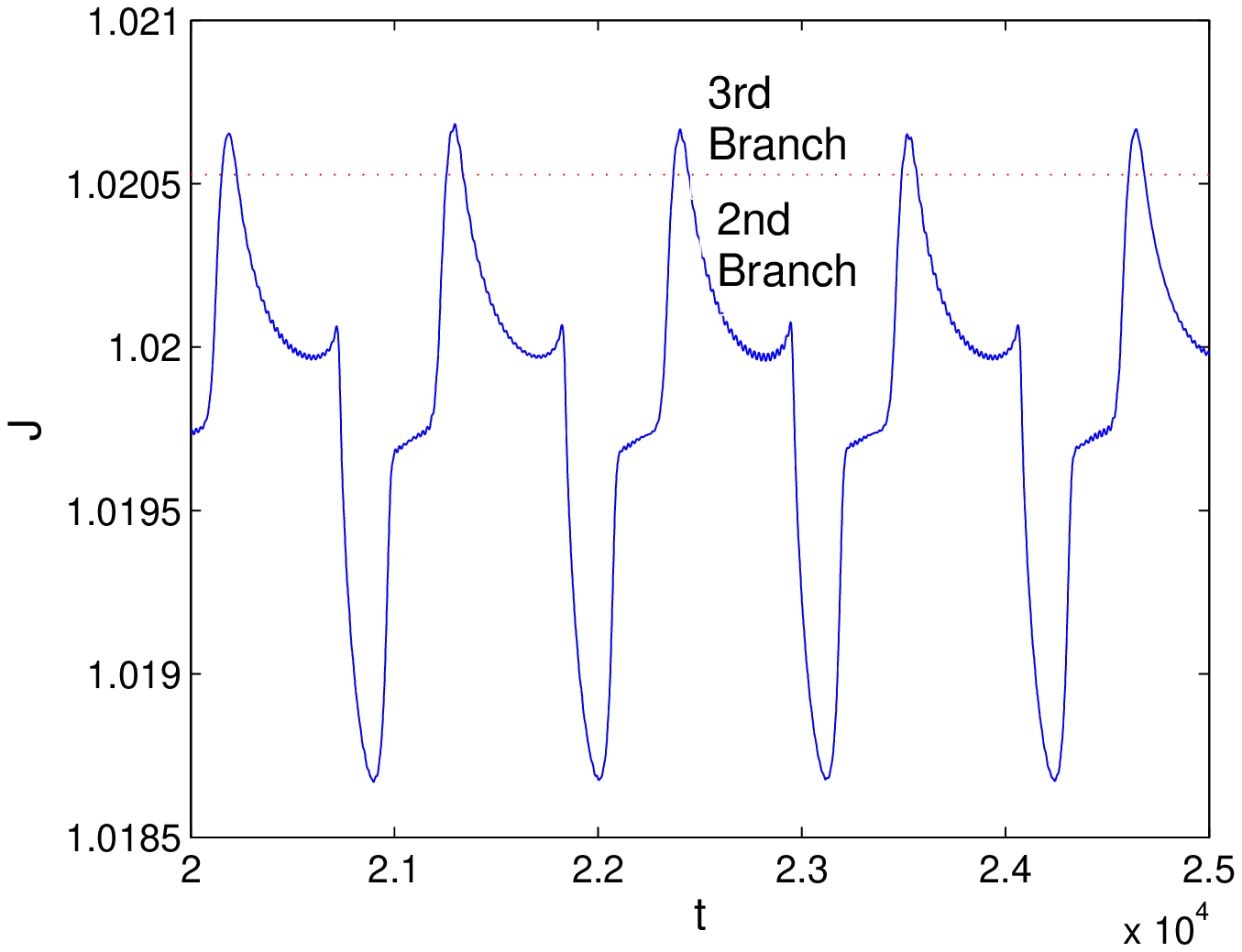}
\includegraphics[width=0.5\linewidth]{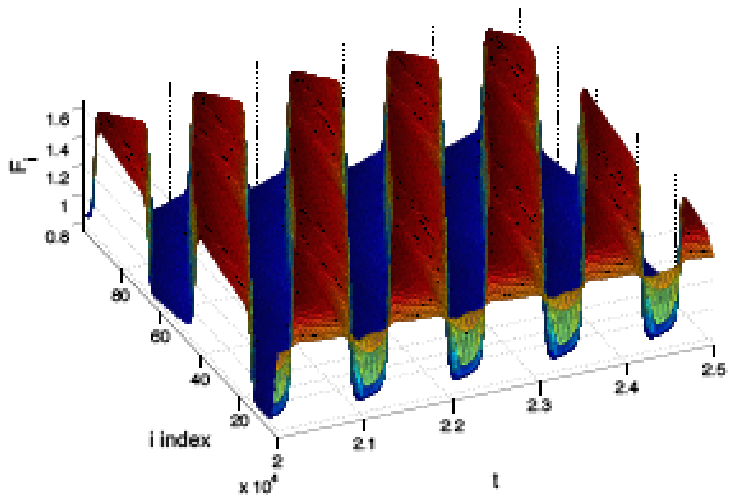}
\caption{(Color online) Current vs time (top) and field distribution vs time (bottom)
displaying dipole-mediated bulk SSCOs for an injecting contact conductivity $\sigma=
1.05231$ and $x=0.25$ (cf. Fig.~\ref{fig2}). 
These oscillations correspond to having a
finite wave train. The other parameter values correspond to
configuration II in Table \ref{t1} and are $N=99$, $\nu= 33.62275$,
$\delta=5.91\times 10^{-4}$, $V=1.22732$, and $\sigma_{2} = 0.73075$.}
\label{fig5}
\end{center}
\end{figure}

\begin{figure}[t!]
\begin{center}\leavevmode
\includegraphics[width=0.5\linewidth]{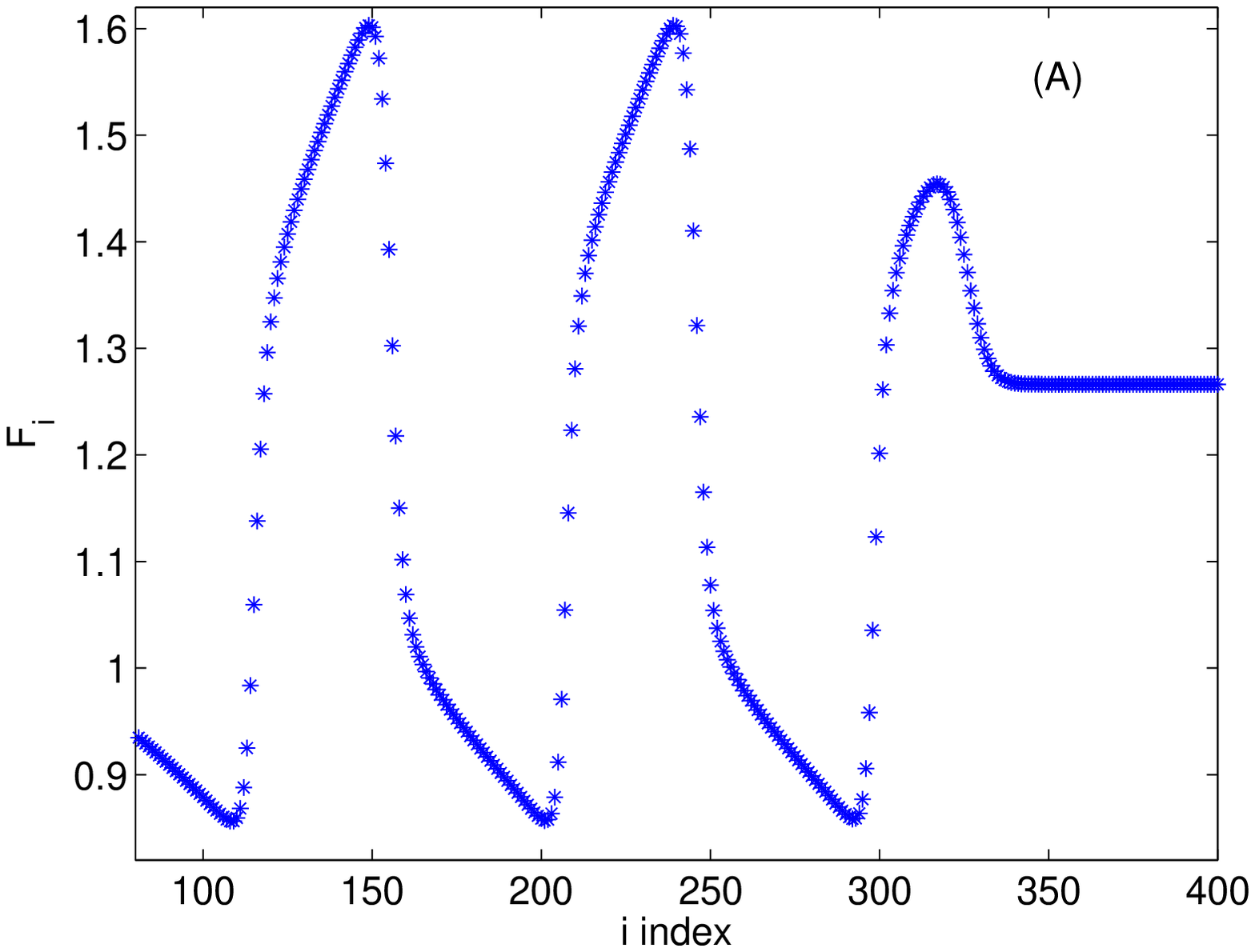}
\includegraphics[width=0.5\linewidth]{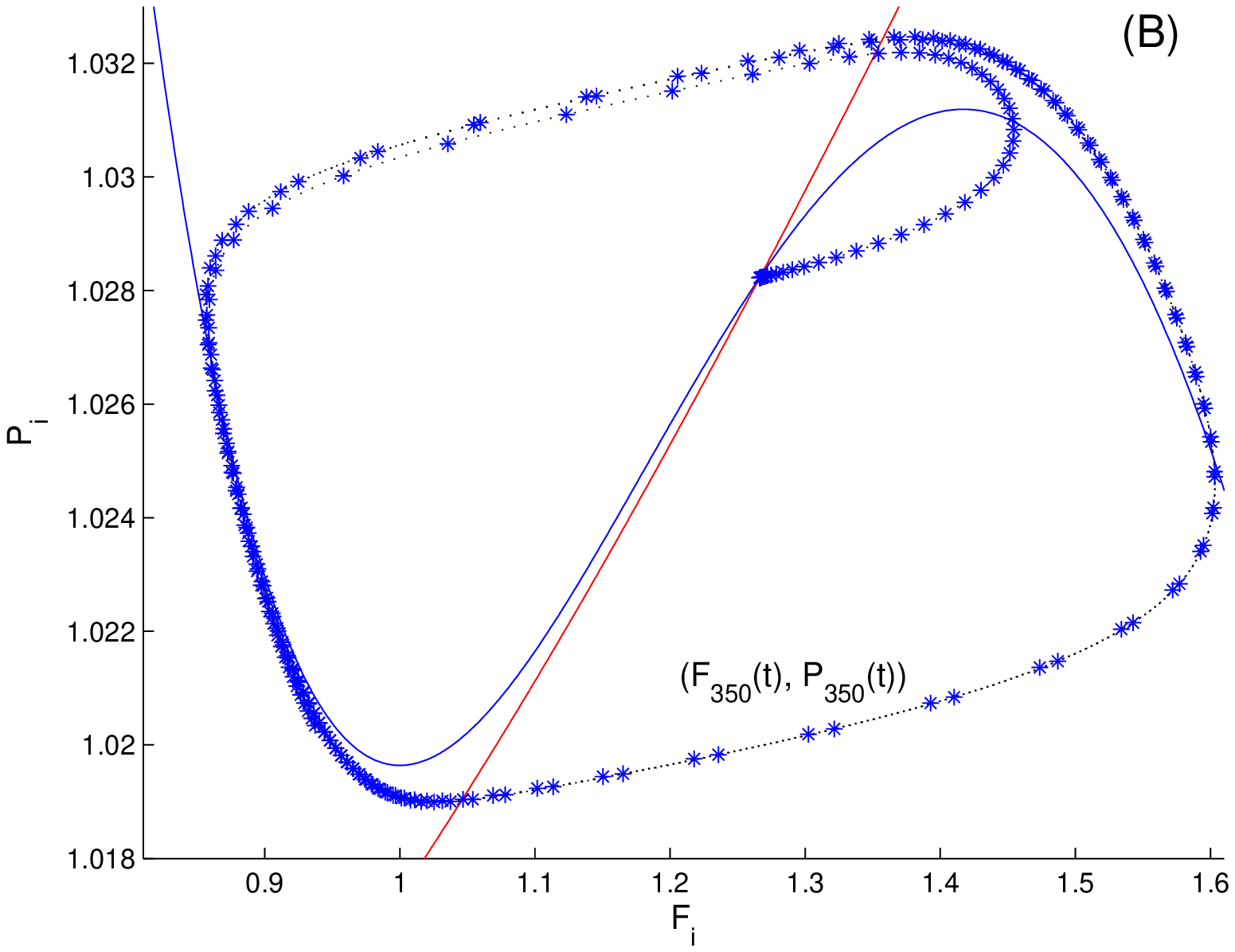}
\caption{(Color online) (a) Numerically obtained field profile of a pulse moving with positive speed
followed by a wave train when there is only one critical point in the phase plane. (b) Phase
plane showing the nullclines and the motion of the 350th QW as the pulse traverses it. The
total current density is $J=1.0165$ and the other parameters correspond to configuration I in
Table \ref{t1}. }
\label{fig6}
\end{center}
\end{figure}

When the conductivity of the injecting contact is such that $\sigma F$ intersects $j(F)$ near
the maximum thereof, it is possible to have SSCOs that are quite different from the ones
appearing in $n$-doped SLs. For an injecting contact conductivity $\sigma=1.05231$ (cf.
Fig.~\ref{fig2}), 
pulses (charge dipoles) may be triggered at the injecting contact,
move toward the receiving contact, and cause SSCOs as shown in Fig.~\ref{fig5}. 
For all
the instantaneous values of $J(t)$ during these SSCOs, there is only one constant solution
of Eq.~(\ref{17}): most of the times this solution is on the second NDR branch
of $j(F)$. Only when $J(t)$ is near its maximum value, the constant solution is on the
third branch of $j(F)$. For a constant current density, the constant solution of
Eq.~(\ref{17}) is on the NDR branch, which implies that the system is oscillatory and
that periodic wave trains are possible. The realization of wave trains for a long dc-biased SL
are displayed in Fig.~\ref{fig6}: at any time during SSOC, there are only two fully 
developed pulses present in this SL. These pulses experience variations in their shape and 
velocity when they are generated or arrive at the contacts, but longer SLs allow for 
realizations of wave trains, in which more pulses exist simultaneously inside the SL.

\begin{figure}[t!]
\begin{center}\leavevmode
\includegraphics[width=0.5\linewidth]{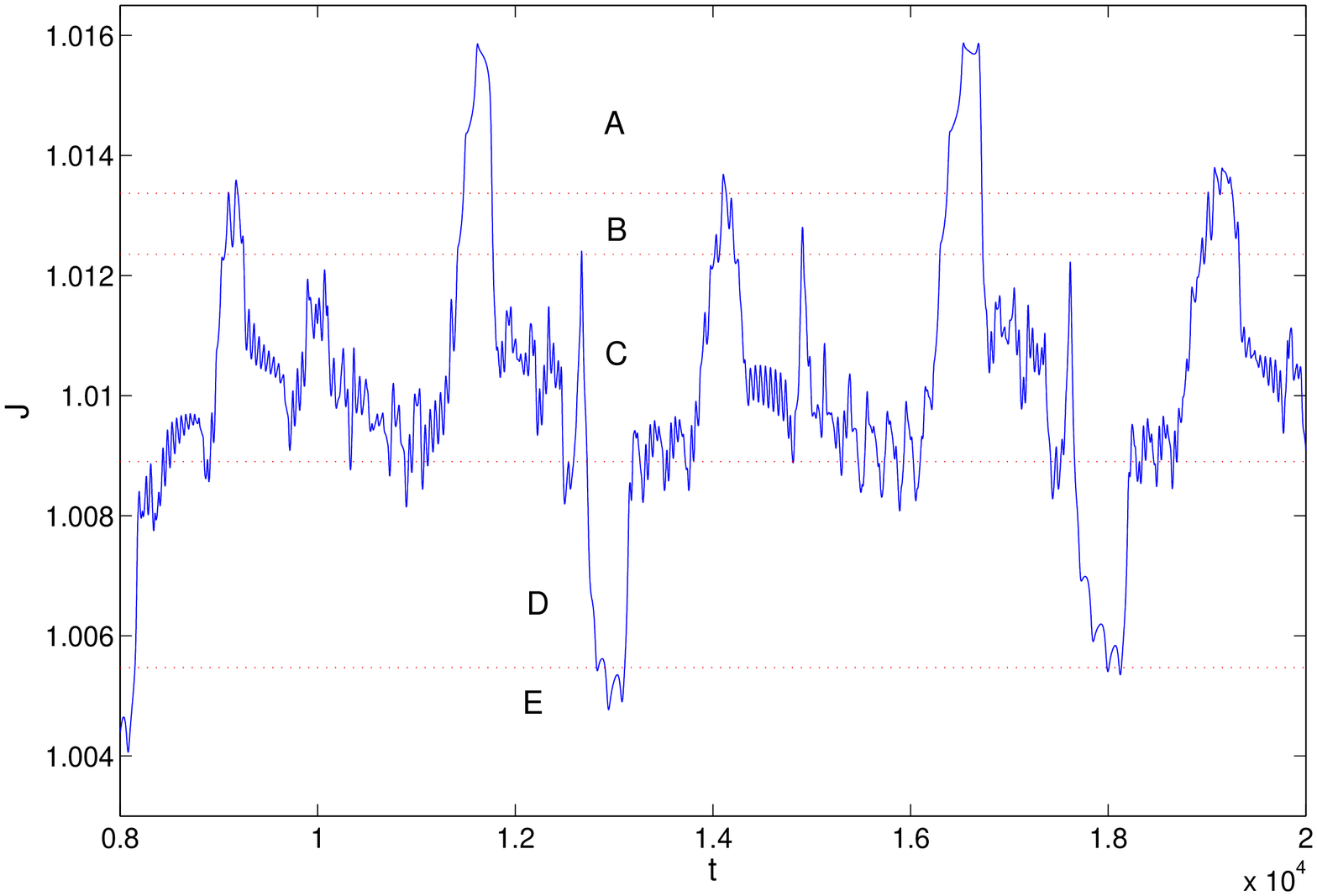}
\includegraphics[width=0.5\linewidth]{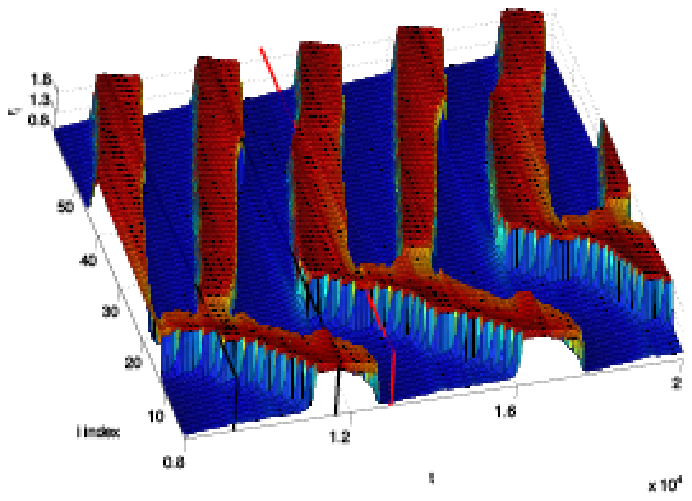}
\caption{(Color online) Current vs time (top) and field distribution vs time (bottom)
displaying dipole-mediated bulk SSCOs for an injecting contact conductivity $\sigma=1.107$
(cf. Fig.~\ref{fig2}). 
The other parameter values correspond to configuration III in Table \ref{t1} and
are $N=61$, $\nu= 70.50677$, $\delta=
4.069\times 10^{-4}$, $V =1.269848$, and $\sigma_{2} = 1.0829727$.}
\label{fig7}
\end{center}
\end{figure}

In the previous example, pulses always move downstream, from left to right. For slightly
larger conductivity of the injecting contact ($\sigma=1.107$ in Fig.~\ref{fig2}), 
Fig.~\ref{fig7} 
shows that two pulses are formed inside the SL and
move with opposite velocities toward the contacts. These pulses are similar to the ones
constructed above for the case of dc current bias (with positive or negative velocity), except
that the current changes slowly with time during the self-oscillation and the pulses
accommodate their form to the instantaneous value of the current. Note that the pulses are
triggered inside the SL, not at the injecting contact. A
pulse moving with positive speed has a long trailing region, in which the field increases as
we move away from the pulse. However, there is a depletion layer near the injecting
contact, in which the field decreases as we move away from the injecting contact. In a long, but
finite SL, a local maximum is formed inside the SL, when the decreasing field near the contact
meets the increasing field in the trailing region of the exiting pulse with positive speed. The
current increases as the exiting pulse is absorbed by the receiving contact, until it surpasses a
critical value. In this case, the local maximum of the field profile inside the SL is split, and two new
pulses are created. The pulse closer to the injecting contact moves toward it with negative
speed whereas the other pulse moves toward the receiving contact with positive speed. The
upward moving pulse reaches the injecting contact and is absorbed there before the downward
moving pulse arrives at the other contact. In this case, the field profile close to the injecting contact
is quasi-stationary, and a local maximum of the field is formed when we match this region
with the trailing region of the downward moving pulse. After the critical value of the current
is reached, another pulse pair is nucleated, and the same process is periodically repeated.

It is interesting to evaluate in some detail the process of nucleation and disappearance of
pulses during these SSCOs. In the current vs time diagram in Fig.~\ref{fig7}, 
we distinguish different regions depending on the number of constant solutions of Eq.~(\ref{17})
that exist for the corresponding instantaneous value of $J(t)$. In region A,
there is only one constant solution on the third branch of $j(F)$. In region B, there is
one constant solution on the third branch and two on the second branch of $j(F)$. In region
C, there are three constant solutions, one on each branch of $j(F)$, whereas two of these
solutions are on the second branch and one of the first branch of $j(F)$, if $J(t)$ is in region
D. There is only one constant solution located on the  first branch of $j(F)$, if $J(t)$ is in region
E. For a constant voltage bias, a pulse moving upstream may be generated only if $J(t)$ surpasses
a critical value (1.007454), which is located in region C and marked by an arrow. Once
generated, the upstream moving pulses persist for any instantaneous value of the current
density. These SSOC are apparently weakly chaotic: we have calculated the
corresponding Lyapunov exponents and found that there is a single positive exponent with
a rather small value of $7.9$$\times$$10^{-6}$.

\begin{figure}[t!]
\begin{center}\leavevmode
\includegraphics[width=0.5\linewidth]{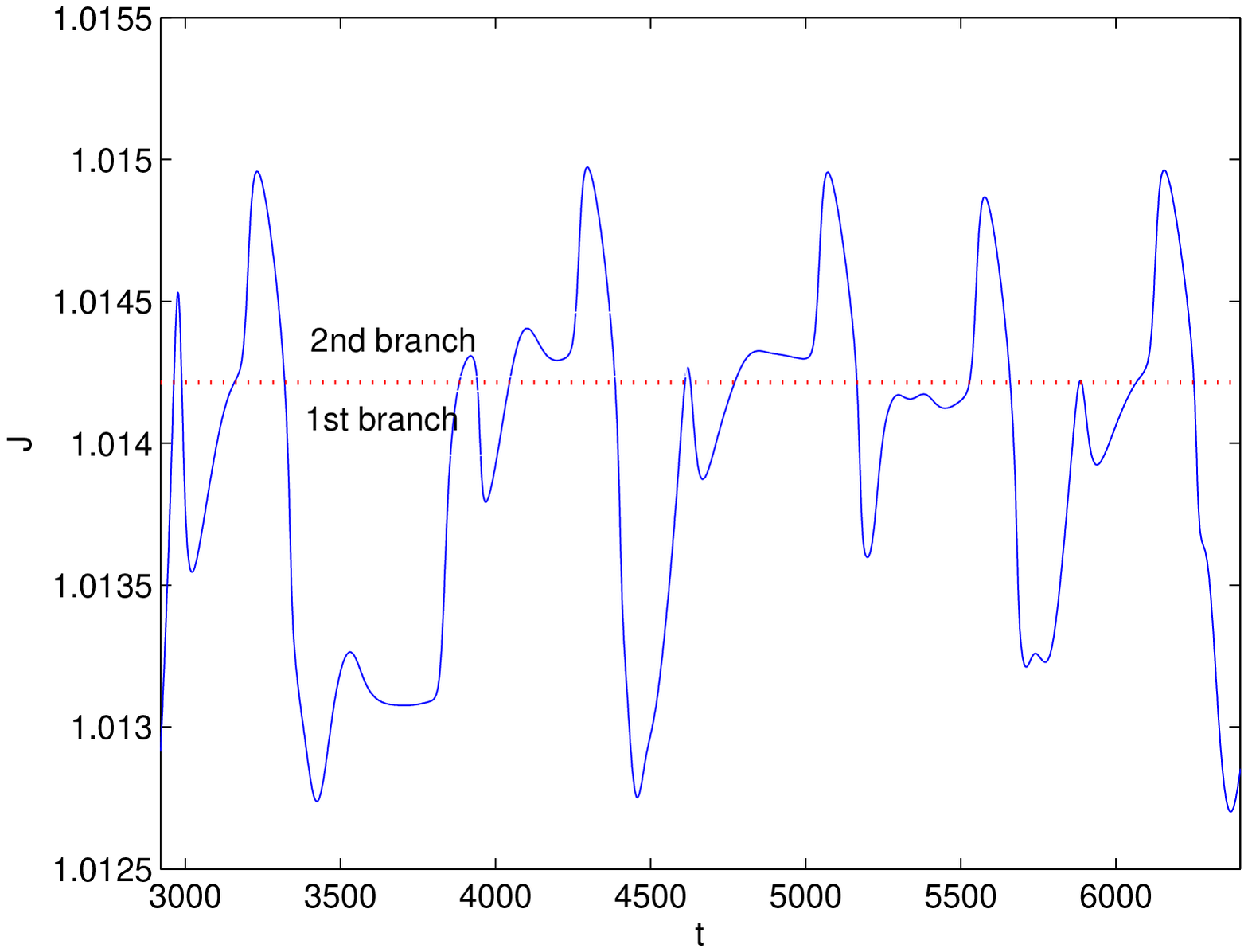}
\includegraphics[width=0.5\linewidth]{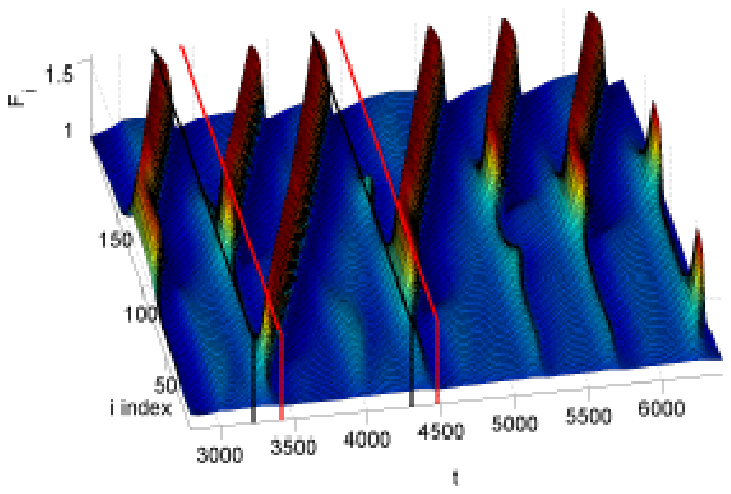}
\caption{(Color online) Current vs time (top) and field distribution vs time (bottom)
displaying dipole-mediated bulk SSCOs for an injecting contact conductivity $\sigma=
1.083425$ (cf. Fig.~\ref{fig2}). 
The other parameter values correspond to configuration I in Table \ref{t1} and are
$N=199$, $\nu= 8.44566$, $\delta=3.7287\times 10^{-3}$, $V=1.00417485$, and $\sigma_{2} = 1.0523129$.}
\label{fig8}
\end{center}
\end{figure}

We have also observed SSCOs mediated by dipole waves that nucleate alternatively at
two different QWs of the SL as shown in Fig.~\ref{fig8}. 
During these SSCOs, there is only
one constant solution of Eq.~(\ref{17}) for any instantaneous value of $J(t)$.
This solution is either on the first or the second branch of $j(F)$. At about $t= 3232$, where
$J(t)$ reaches its global maximum, two dipole waves are nucleated at two different QWs.
They become fully developed pulses and move with positive speed toward the receiving
contact. When the first one arrives there, the current increases so that the corresponding
constant solution of Eq.~(\ref{17}) is on the second branch of $j(F)$. In this case, a
small dipole wave is nucleated at the local field maximum, where the tail of the first dipole
meets the depletion layer near the injecting contact. The small dipole wave never grows into
a fully developed field pulse, and it continues advancing, until the old large dipole wave
disappears at the receiving contact. At this time (about 4296), a large current spike appears,
and a new dipole wave is formed closer to the injecting contact than the small dipole. $J(t)$
decreases abruptly, while the small dipole disappears, and the newly created dipole reaches
a large size and moves toward the receiving contact. Shortly afterward, a new current spike
marks the creation of another small dipole. This small dipole travels toward the receiving
contact, and it grows only when the only existing large pulse reaches the receiving contact and
disappears. A small dipole formed closer to the injecting contact does not grow, until the
large pulse reaches the receiving contact and disappears without triggering a new dipole wave.
In this case, the corresponding pulse is close to the receiving contact. When it reaches the contact
and disappears into it, two new dipoles are simultaneously triggered and become fully developed.
A scenario similar to the one previously described follows, marked again by a large current spike.
The situation is not repeated exactly: there are small differences in the QWs at which pulses are nucleated,
differences in the size and lifetimes of the small dipoles, etc. These oscillations also seem
to be weakly chaotic, in which a single Lyapunov exponent is small and positive.

If the injecting contact conductivity is smaller so that the contact current $\sigma F$ intersects
$j(F)$ on the NDR branch thereof, there appear standard SSCOs due to
repeated dipole pulse nucleation at the injecting contact and motion toward the receiving contact.

\section{Conclusions}
We have calculated the recombination coefficient as a function of the applied
electric field for undoped, photo-excited, weakly coupled GaAs/Al$_{x}$Ga$_{1-x}$As superlattices.
Depending on the Al content $x$, the superlattice may have only one static domain for
small $x$ or two stable differentiated static domains for $0.45<x\leq 1$. In the latter case,
a dc voltage biased SL under weak photoexcitation may exhibit self-sustained oscillations of 
the current due to repeated nucleation of a charge monopole or dipole waves at the injecting 
contact and their motion toward the collector. For high photo-excitation intensities, the walls 
separating electric field domains are mostly pinned, and self-sustained oscillations of the 
current occur only in narrow voltage intervals. For small $x$ among other unusual 
phenomena, there may appear weakly chaotic SSOC due to dipole dynamics in 
dc-voltage-biased SLs for
high photo-excitation intensities, during which nucleated dipole waves can split in two oppositely moving dipoles.
These dipoles are pulses of the electric field with shapes and behavior similar to pulses in
excitable media, where a sufficiently large disturbance about the unique stable domain may
induce them. For other parameter values, the unique static domain is unstable, and the
underlying dynamics is oscillatory so that wave trains formed by succession of pulses give rise
to self-sustained oscillations of the current.

\section*{Acknowledgements}
The work was financially supported in part by the Spanish Ministry of Science and
Innovation under grant FIS2008-04921-C02-01.

\end{document}